\documentclass[preprint,nofootinbib,superscriptaddress]{revtex4}
\usepackage[dvips]{graphicx}
\renewcommand{\slash}{\displaystyle{\not}}
\newcommand{\beq}{\begin{equation}}
\newcommand{\eeq}{\end{equation}}
\newcommand{\bea}{\begin{eqnarray}}
\newcommand{\eea}{\end{eqnarray}}

\begin{document}
\preprint{
 {\vbox{
 \hbox{MADPH--06--1466}
 \hbox{hep-ph/0608215}}}}
 
\title{Neutralino annihilation to $q \bar q g$}

\author{Vernon~Barger}
\email{barger@pheno.physics.wisc.edu}
\affiliation{Department of Physics, University of Wisconsin, 1150 University
Avenue, Madison, Wisconsin 53706 USA}

\author{Wai-Yee~Keung}
\affiliation{Physics Department, University of Illinois at Chicago, 
Illinois 60607--7059 USA}

\author{Heather~E.~Logan}
\email{logan@physics.carleton.ca}
\affiliation{Ottawa-Carleton Institute for Physics, 
Carleton University, Ottawa K1S 5B6 Canada}

\author{Gabe~Shaughnessy}
\email{gshau@physics.wisc.edu}
\affiliation{Department of Physics, University of Wisconsin, 1150 University
Avenue, Madison, Wisconsin 53706 USA}

\begin{abstract}
We compute the cross section for $\chi\chi \to q \bar q g$ at order
$\alpha_s^2/M_{\widetilde q}^6$ arising from interference between the 
tree-level and loop-induced processes.  This interference term 
is the same order
in $\alpha_s$ as $\chi\chi \to gg$; for mass degenerate squarks
$M_{\widetilde q_R} = M_{\widetilde q_L} = M_{\widetilde q}$ we find
$v_{\rm rel} \sigma_{\rm int} = 
[-2 m_{\chi}^2/3 M_{\widetilde q}^2] \, v_{\rm rel} \sigma (\chi\chi \to gg)$.
\end{abstract}

 
\maketitle
\section{Introduction}

The presence of non-baryonic dark matter in the universe is compelling
evidence for physics beyond the Standard Model.  
Supersymmetry (SUSY) provides an especially attractive explanation with 
the lightest supersymmetric particle (LSP), usually a neutralino $\chi$,
as the dark matter candidate; for a recent review, see Ref.~\cite{dmrev}.

\begin{figure}[t]
\includegraphics*{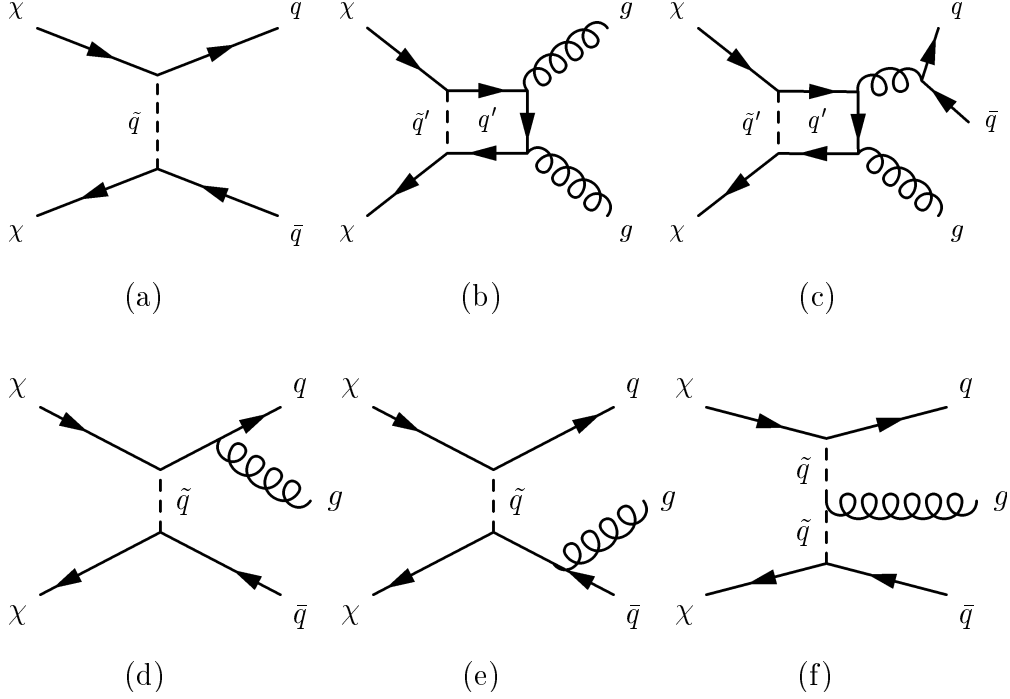}
\caption{Feynman diagrams that contribute to neutralino annihilation to
quarks/gluons.  Shown are the tree-level diagram for 
$\chi\chi \to q\bar q$ (a), typical one-loop diagrams for $\chi\chi\to gg$ (b)
and $\chi \chi \to q \bar q g$ (c), and the three tree-level diagrams 
contributing to the dimension-eight operator for $\chi\chi \to q \bar q g$
(d-f).
}
\label{fig:fd}
\end{figure}

We focus in this paper on the QCD corrections to neutralino annihilation.
We therefore consider only the annihilation processes involving 
quarks and/or gluons in the final state.  We also assume that the LSP
is largely gaugino, as is often the case in mSUGRA models~\cite{msugra}; in 
this case neutralino annihilation via $s$-channel Higgs or $Z$ exchange
is suppressed, since these particles require a Higgsino admixture to 
couple to the LSP.  We thus focus on processes that involve an internal squark
exchange, as shown in Fig.~\ref{fig:fd}.

The leading contribution to neutralino annihilation via exchange of
a squark of mass $M_{\widetilde q}$, shown in Fig.~\ref{fig:fd}(a),
can be reduced to an effective vertex described by a dimension-six operator
suppressed by $M_{\widetilde q}^2$,
\beq
  \mathcal{L}=(c/M_{\widetilde q}^2) \mathcal{O}_6~,
  \hspace{1cm} 
  \mathcal{O}_6=(\bar \chi \gamma_\mu \gamma_5 \chi )
   (\bar q \gamma^\mu \gamma_5 q),
\eeq
where the dimensionless constant $c$ contains the relevant couplings of the 
process.  If the annihilation occurs at rest, the neutralino spinors reduce to 
a spin-singlet combination,
$\bar \chi \gamma_\mu \gamma_5 \chi \rightarrow 
K_\mu \bar \chi (\gamma_5 /\sqrt{2} m_\chi) \chi$, 
where $K$ is the center-of-mass
momentum of the two-neutralino system.  This result can be understood by 
considering that the Majorana nature of neutralinos makes the initial state 
behave as an effective pseudoscalar (details are given in 
Sec.~\ref{sec:calc}).  The operator ${\cal O}_6$ can then be written as the 
divergence of the axial-vector current,
\beq
 {\cal O}_6 \to 
 \left[ \overline\chi \ (i\gamma_5 / \sqrt{2}m_\chi)  \chi \right]  \,  
 \left[\partial_\mu (\overline q \gamma^\mu\gamma_5  q) \right]\, .
\eeq
In the massless quark limit, $m_q=0$, the axial vector current is conserved 
at tree level, $\partial_\mu (\overline q \gamma^\mu\gamma_5  q)=0$, and
all tree-level dimension-six amplitudes vanish in this limit.  
This is the well known partially-conserved axial current (PCAC) condition.

The quark mass suppression can be lifted in two ways:
\begin{enumerate}
\item by going beyond leading order in $\alpha_s$ to include the correction 
to the dimension-six operator that involves the anomalous triangle diagram;
\item by going to dimension-eight by including hard gluon radiation.
\end{enumerate}

The anomalous triangle diagram is shown in Fig.~\ref{fig:fd}(b) (the 
quark triangle appears at dimension-six when the heavy squark line is 
shrunk to a point).  This diagram contributes to $\chi\chi \to gg$
and yields a cross section parametrically of order 
$\alpha_s^2 m_{\chi}^2/M_{\widetilde q}^4$.  The analogous process
$\chi\chi \to \gamma\gamma$ was first studied using the anomaly 
equation in Refs.~\cite{Rudaz:1989ij,Bergstrom}; a modern computation of 
$\chi\chi \to gg$ in the $v_{\rm rel} = 0$ limit with arbitrary 
neutralino composition and squark mixing, and keeping $m_{q^{\prime}} \neq 0$ 
in the loop, was performed in Ref.~\cite{Dreesgg}. 
Ref.~\cite{Gounaris} extended the calculation to arbitrary $v_{\rm rel}$,
with results presented in a set of numerical codes~\cite{platon}.

The next-to-leading order (NLO) QCD corrections to $\chi\chi\to gg$ were
recently computed in Ref.~\cite{Barger05} in the approximation 
$m_{q^{\prime}} = 0$ and $M_{\widetilde q} \gg m_{\chi}$.  These corrections
are of order $\alpha_s^3 m_{\chi}^2/M_{\widetilde q}^4$,
and provide about a 60\% enhancement over the leading order (LO) 
$\chi\chi \to gg$ cross section for typical parameter values~\cite{Barger05}.

The tree-level process $\chi\chi \to q \bar q g$ corresponds in the limit 
$m_q = 0$ to a dimension-eight operator, with the amplitude suppressed
by $1/M_{\widetilde q}^4$.  The relevant diagrams are shown in 
Fig.~\ref{fig:fd}(d-f).  The similar process 
$\chi\chi \to f \bar f \gamma$ was calculated in Refs.~\cite{Bergstrom,FOR}
for a photino LSP.
An exact tree-level calculation of $\chi\chi \to q \bar q g$
with arbitrary neutralino composition and squark mixing,
for $m_q = 0$ and $v_{\rm rel} = 0$, was done in 
Ref.~\cite{Dreesgg} and has a resulting cross section  of order
$\alpha_s m_{\chi}^6/M_{\widetilde q}^8$.

In this paper we compute the interference term between the tree-level
$\chi\chi \to q \bar q g$ diagrams of Figs.~\ref{fig:fd}(d,e,f) 
and the one-loop $\chi\chi\to q \bar q g$
process of Fig.~\ref{fig:fd}(c).  This interference term is 
parametrically of order $\alpha_s^2 m_{\chi}^4/M_{\widetilde q}^6$.
Counting $\alpha_s$ and $m_{\chi}^2/M_{\widetilde q}^2$ as comparable
suppression factors, this interference term is ``the same order'' as 
both the dimension-eight process and the NLO QCD corrections to 
$\chi\chi \to gg$, as summarized in Table~\ref{tab:literature}.
Counting only the order in $\alpha_s$, the interference term is the same 
order as $\chi \chi \to gg$.

\begin{table}
\begin{tabular}{lccc}
\hline\hline
Process & Order & Ref. & Approximation \\
\hline
$\chi\chi \to gg$, Dim-6 &
  $\alpha_s^2 m_{\chi}^2/M_{\widetilde q}^4$ &
  \cite{Dreesgg} &
  exact for $v_{\rm rel} = 0$ \\
\hline
$\chi\chi \to q \bar q g$, Dim-8 &
  $\alpha_s m_{\chi}^6/M_{\widetilde q}^8$ &
  \cite{Dreesgg} &
  exact for $m_q = 0$ \\
$\chi\chi \to q \bar q g$, Dim-6/8 interference &
  $\alpha_s^2 m_{\chi}^4/M_{\widetilde q}^6$ &
  [this paper] &
  leading $1/M_{\widetilde q}^6$ term, $m_{q,q^{\prime}} = 0$ \\
$\chi\chi \to gg$ at NLO, Dim-6 &
  $\alpha_s^3 m_{\chi}^2/M_{\widetilde q}^4$ &
  \cite{Barger05} &
  leading $1/M_{\widetilde q}^4$ term, $m_{q^{\prime}} = 0$ \\
\hline\hline
\end{tabular}
\caption{Relevant neutralino annihilation calculations and their 
relative suppression by powers of $\alpha_s$ and 
$m_{\chi}^2/M_{\widetilde q}^2$.}
\label{tab:literature}
\end{table}

Throughout this paper we assume that the LSP is lighter than the top 
quark, so that $\chi\chi \to t \bar t$ is kinematically inaccessible.  
(The QCD corrections to $\chi\chi \to Z^*, h^* \to t \bar t$ near 
threshold were recently computed in Ref.~\cite{Moroi}.)
We take the five light quark species to be massless.
As noted earlier, we also assume that the neutralinos annihilate at rest and that 
the LSP is gaugino-like (the coupling of the Higgsino component
to quark-squark pairs is proportional to the quark mass, taken here
to be zero).
We also assume that the squarks are heavy compared to the LSP,
$M_{\widetilde q} \gg m_{\chi}$, and work to leading order in the expansion
in $m_{\chi}^2/M_{\widetilde q}^2$.
In the calculation of the quark/squark loop in Fig.~\ref{fig:fd}(c) 
we sum over five massless internal quark species, and ignore the top
quark contribution; the massive quark loop decouples very quickly
with decreasing $m_{\chi}/m_t$.

\section{Calculation}
\label{sec:calc}

In the limit of zero relative neutralino velocity, $v_{\rm rel} = 0$, a 
neutralino pair behaves as a pseudoscalar due to the Majorana nature of the
particles.  In particular, the antisymmetrized wave function of the 
initial-state identical particles $\chi\chi$, of total momentum $K$, can
be reduced to a projection operator for each of the four spin combinations 
of the initial-state neutralinos:
\begin{eqnarray}
     u_1(K/2)\bar v_2(K/2)- u_2(K/2) \bar v_1(K/2) 
   = (m_\chi+\slash K/2)\gamma_5
   \qquad &{\rm for}& \quad
  | \rightarrow \rangle | \leftarrow \rangle
  \quad {\rm or} \quad
  - | \leftarrow \rangle | \rightarrow \rangle,
  \nonumber \\
     u_1(K/2)\bar v_2(K/2)- u_2(K/2) \bar v_1(K/2) 
   = 0
     \qquad &{\rm for}& \quad
  | \rightarrow \rangle | \rightarrow \rangle
  \quad {\rm or} \quad
  | \leftarrow \rangle | \leftarrow \rangle.
\end{eqnarray}
The cross section is then obtained by averaging over the squares of the amplitudes of the two contributing spin states.

One can alternatively use the spin singlet initial state,
\begin{equation}
  \left( | \rightarrow \rangle | \leftarrow \rangle
  - | \leftarrow \rangle | \rightarrow \rangle \right)/\sqrt{2},
\end{equation}
which produces an additional $\sqrt{2}$ in the antisymmetrized spinor
combination and in the resulting matrix elements:
\begin{equation}
     u_1(K/2)\bar v_2(K/2)- u_2(K/2) \bar v_1(K/2) 
   = \sqrt{2} \left( m_\chi + \slash K/2 \right) \gamma_5
\qquad \qquad ({\rm spin \ singlet}).
   \label{eq:uv-uv}
\end{equation}
Either approach leads to the same overall result for the annihilation cross 
section; we use Eq.~(\ref{eq:uv-uv}) with the additional $\sqrt{2}$ 
in what follows in order to obtain the matrix element for the spin-singlet 
state.
(After squaring the matrix element, one must still average over the four 
initial neutralino spin states, yielding a factor of $1/4$ in the cross 
section.)

\subsection{Dimension-six amplitude}
\label{sec:dim6}

In the $m_q = 0$ limit, the leading contribution to the dimension-six
amplitude arises at order $\alpha_s^2$ from the loop-induced process 
$\chi\chi \to gg$.  In the language of the partially conserved axial
current, this can be expressed as the anomaly equation~\cite{adler},
\beq
   \partial_\mu (\bar q \gamma^\mu \gamma_5 q) 
   = 2 m_q \bar q i \gamma_5 q 
   + \frac{\alpha_s}{4\pi} G_{\mu \nu} ^{(a)} \widetilde G^{(a) \mu \nu},
   \label{eq:anomaly}
\eeq
where 
$\frac{1}{2} \widetilde G_{\mu \nu} 
= \epsilon_{\mu \nu \alpha \beta} G^{\alpha \beta}$ 
is the dual of the color field strength tensor.  For $m_q \ll m_\chi$, 
the first term on the right-hand side of Eq.~(\ref{eq:anomaly}) can
be neglected; the dimension-six amplitude can then be expressed in the
form
\beq
    {\cal M}_{\rm loop}(\chi \chi \to g g) 
    = \left(\frac{c/m_\chi}{2 M_{\widetilde q}^2}\right) 
    (\bar \chi i \gamma_5 \chi) 
    \frac{\alpha_s}{4 \pi}G_{\mu \nu} ^{(a)} \widetilde G^{(a) \mu \nu}.
\eeq
The Feynman diagrams for $\chi\chi \to gg$ consist of Fig.~\ref{fig:fd}(b),
the analogous diagram with one neutralino and one gluon line interchanged,
and the corresponding diagrams with the gluons or the neutralinos crossed.
The amplitude for $\chi\chi \to gg$ in the massless quark limit is 
given by
\beq
    {\cal M}_{\rm loop} = \frac{\alpha_s}{2\sqrt{2} \pi}
    \sum_{q^{\prime}} \left[ \frac{|g_r|^2}{M_{\widetilde q_R^{\prime}}^2}
      + \frac{|g_{\ell}|^2}{M_{\widetilde q_L^{\prime}}^2} \right]
    i \epsilon^{\mu\nu\alpha\beta} \epsilon^{*a}_{\mu}(k_1) 
    \epsilon^{*a}_{\nu}(k_2) k_{1\alpha} k_{2\beta},
    \label{eq:Mloop}
\eeq
where the sum runs over the five light (massless) quark flavors $q^{\prime}$. 
An expression for the expansion of $\mathcal{M}_{\rm loop}$ in powers
of $m_{\chi}^2/M_{\widetilde q}^2$ is given in the appendix.
The $\chi q^{\prime} \widetilde q^{\prime}_{R,L}$ 
couplings $g_{r,\ell}$ for right and left quark helicities, respectively, are
\begin{equation}
  g_r = -\sqrt 2 N_{11} g^{\prime} Q, \hspace{0.5in} 
  g_{\ell} = - \sqrt 2 N_{11} g^{\prime} (T_3 - Q) + \sqrt 2 N_{12} g T_3.
  \label{eq:grgl}
\end{equation}
Here $T_3$ is the quark isospin, $Q$ is the quark electric charge, 
$g$ and $g^{\prime}$ are the weak couplings under $SU(2)_L$ and 
$U(1)_Y$, respectively, 
and $N_{11}$ and $N_{12}$ are the bino and wino components of the 
neutralino as defined in Ref.~\cite{DarkSUSY}. 
The matrix element in Eq.~(\ref{eq:Mloop}) results in an annihilation cross 
section for $\chi \chi \to gg$ of 
\begin{equation}
  v_{\rm rel} \sigma_{\rm dim6} = \frac{\alpha_s^2}{32 \pi^3} m_{\chi}^2
  \left\{ \sum_{q^{\prime}} 
  \left[ \frac{|g_r|^2}{M_{\widetilde q^{\prime}_R}^2}
    + \frac{|g_{\ell}|^2}{M_{\widetilde q^{\prime}_L}^2} \right] \right\}^2.
\label{eq:dim6xsec}
\end{equation}

For massless quarks in the loop, taking one of the final-state
gluons off-shell by an amount $q^2 \neq 0$ yields a relative shift in the 
loop integral by $q^2/M_{\widetilde q}^2$; in particular, this shift is 
higher order in the $1/M_{\widetilde q}^2$ expansion than the leading term. 
To leading $1/M_{\widetilde q}^2$ order it is
then straightforward to extend the loop amplitude to include
one off-shell gluon splitting into a quark-antiquark pair, shown 
schematically in Fig.~\ref{fig:fd}(c).  The matrix element is
\beq
    {\cal M}_{\rm split} = 
    - \frac{g_s}{\sqrt{2}} 
    \sum_{q^{\prime}} \left[ \frac{|g_r|^2}{M_{\widetilde q_R^{\prime}}^2}
      + \frac{|g_{\ell}|^2}{M_{\widetilde q_L^{\prime}}^2} \right]
    i \epsilon^{\mu\nu\alpha\beta} \epsilon^{*c}_{\mu} q_{3\alpha}
    (q_1 + q_2)_{\beta}
    \frac{\alpha_s}{2 \pi (q_1 + q_2)^2}
    \bar u(q_1) T^c \gamma_{\nu} v(q_2),
\label{eq:dim6}
\eeq
where $q_1$, $q_2$, and $q_3$ are the momenta of the final-state quark,
antiquark, and gluon, respectively, and the sum runs over the five 
light (massless) quarks in the loop.  Note that the gluon splitting couples 
the box diagram to a \emph{vectorlike} quark current.

The genuine anomaly loop is a closed fermion loop which has
a negative sign. The loop in the present case is basically a box with
one side as the scalar quark line, so it is not a closed fermion
loop. However, it does have a relative negative sign with respect to
the dimension-eight amplitude (see the next subsection) 
because the two fermion lines are topologically
twisted -- replacing the neutralino spinor pair with the zero-velocity
result from Eq.~(\ref{eq:uv-uv}) requires contracting the two spinors
together, yielding an extra overall minus sign.  
The situation is similar to that of the two amplitudes in
Bhabha scattering.

Using Eq.~(\ref{eq:dim6}) we can compute the zero-velocity neutralino 
annihilation cross section from the gluon-splitting process.  Squaring
the matrix element, summing over final-state polarizations and colors and
averaging over initial-state polarizations, we obtain
\begin{equation}
  \frac{1}{4} \sum_{\rm pols} |\mathcal{M}_{\rm split} |^2 = 
  \frac{\alpha_s^3}{\pi} \left\{ \sum_{q^{\prime}}
  \left[ \frac{|g_r|^2}{M_{\widetilde q^{\prime}_R}^2} 
      + \frac{|g_{\ell}^2|}{M_{\widetilde q^{\prime}_L}^2} \right] \right\}^2
  \frac{(q_1 \cdot q_3)^2 + (q_2 \cdot q_3)^2}{q_1 \cdot q_2},
\label{eq:msplitsq}
\end{equation}
for a single (massless) final-state quark flavor.

Note the divergence in Eq.~(\ref{eq:msplitsq}) 
as the gluon propagator goes on shell, 
$(q_1 + q_2)^2 = 2 q_1 \cdot q_2 \rightarrow 0$.  
This is the usual 
divergence that appears in NLO calculations when a final-state 
parton is soft or collinear; in this case the divergence is cut off by the 
quark mass and gives rise to a $\log (m_{\chi}^2/m_q^2)$ enhanced term
in the total cross section~\cite{FOR}.  
This logarithmic term is precisely canceled
by the renormalization of the strong coupling due to the quark bubble
that appears in the virtual part of the NLO correction to 
$\chi\chi \to gg$~\cite{Barger05}.  
This is the familiar cancellation of logarithmic
divergences guaranteed by the Kinoshita-Lee-Nauenberg theorem~\cite{KLN}.

\subsection{Dimension-eight amplitude}

The dimension-eight amplitude comes from computing the tree-level 
process $\chi\chi \to q \bar q g$ shown in Fig.~\ref{fig:fd}(d-f)
in the massless quark limit, $m_q = 0$.  Because the $\chi \chi \to q \bar q$
amplitude [Fig.~\ref{fig:fd}(a)] is zero in this limit, the gluon radiation
diagram contains no soft or collinear divergences.
The tree-level matrix element has the general form,
\begin{equation}
  i{\cal M}_{\rm tree}=(i)^5 g_s |g_{r,\ell}|^2
  \bar u(q_1) T^c A^c P_{R,L} v(q_2),
\end{equation}
where the amplitude $A^c$ is given below.  The $i$ in front of $\mathcal{M}_{\rm tree}$ follows the usual Feynman rule
convention. The factor $(i)^5$ counts three vertices and two propagators.
The $SU(3)$ matrix $T^c$ links the outgoing gluon with color label $c$ 
(contained inside $A^c$) to the $q\bar q$ system.  The chirality projection
operators $P_{R,L} = (1 \pm \gamma_5)/2$ reflect the fact that the squarks
couple to quarks of specific helicity.
The pieces of the amplitude are $A^c = A^c_1 + A^c_2 + A^c_3$, with
\begin{eqnarray}
  A^c_1 &=& \frac{- \slash \epsilon^{*c} \frac{1}{\slash q_1 + \slash q_3}
    \frac{\slash K}{\sqrt{2}} \gamma_5}
  {(K/2 - q_2)^2 - M_{\widetilde q_{R,L}}^2}
  \rightarrow
  \frac{- \slash \epsilon^{*c} \gamma_5/\sqrt{2}}
  {(K/2 - q_2)^2 - M_{\widetilde q_{R,L}}^2},
  \nonumber \\
  A^c_2 &=& \frac{- \frac{\slash K}{\sqrt{2}} \gamma_5 
    \frac{1}{-\slash q_2 - \slash q_3} \slash \epsilon^{*c}}
   {(-K/2 + q_1)^2 - M_{\widetilde q_{R,L}}^2}
   \rightarrow
  \frac{+ \slash \epsilon^{*c} \gamma_5/\sqrt{2}}
  {(-K/2 + q_1)^2 - M_{\widetilde q_{R,L}}^2},
  \nonumber \\
  A^c_3 &=& \frac{- \frac{\slash K}{\sqrt{2}} \gamma_5 \ \ \ 
    (q_1 - q_2) \cdot \epsilon^{*c}}
  {\left[ (-K/2 + q_1)^2 - M_{\widetilde q_{R,L}}^2 \right]
    \left[ (K/2 - q_2)^2 - M_{\widetilde q_{R,L}}^2 \right]},
\end{eqnarray}
where $A^c_1$, $A^c_2$, and $A^c_3$ correspond to cases that the gluon is 
radiated from the quark line, the antiquark line and the squark line, 
as in Figs.~\ref{fig:fd}(d,e,f), respectively.
The outgoing four-momenta of the quark, antiquark and gluon are denoted
$q_1$, $q_2$ and $q_3$, respectively, and $K$ is the incoming four-momentum 
of the initial two-neutralino system.
The Dirac equation was applied to $A^c_1$ and $A^c_2$ in the massless
quark limit.  Combining these three contributions, we get
\begin{equation}
  A^c = -\frac{1}{\sqrt{2}} 
  \frac{(q_2 - q_1) \cdot K \, \slash \epsilon^{*c} 
    + (q_1 - q_2) \cdot \epsilon^{*c} \, \slash K} 
  {\left[ (-K/2 + q_1)^2 - M_{\widetilde q_{R,L}}^2 \right]
    \left[ (K/2 - q_2)^2 - M_{\widetilde q_{R,L}}^2 \right]} \gamma_5.
\end{equation}
Note that the pieces with only one squark propagator in the denominator
have canceled in the massless quark limit.
The resulting matrix element is
\begin{equation}
  \mathcal{M}_{\rm tree} = - \frac{g_s}{\sqrt{2}} 
  \left| g_{r,\ell} \right|^2 \bar u(q_1) T^c
  \frac{ (q_2 - q_1) \cdot q_3 \, \slash \epsilon^{*c} 
    + (q_1 - q_2) \cdot \epsilon^{*c} \, \slash q_3}
  {\left[ (-K/2 + q_1)^2 - M_{\widetilde q_{R,L}}^2 \right]
    \left[ (K/2 - q_2)^2 - M_{\widetilde q_{R,L}}^2 \right]}
  \gamma_5 P_{R,L} v(q_2),
\end{equation}
for the diagrams involving $\widetilde q_R$ and $\widetilde q_L$, 
respectively.

Using the Chisholm identity, 
\begin{equation}
  g^{\alpha\beta} \gamma^{\mu} - g^{\alpha \mu} \gamma^{\beta}
  = \gamma^{\alpha} \gamma^{\beta} \gamma^{\mu} - g^{\beta \mu} \gamma^{\alpha}
  - i \epsilon^{\mu \nu \alpha \beta} \gamma_{\nu} \gamma_5,
  \qquad {\rm with} \qquad
  \epsilon^{0123} = +1,
\end{equation}
together with the Dirac equation and the polarization condition
$\epsilon^*_{\mu} q_3^{\mu} = 0$ for the external gluon,
we obtain an amplitude which has a 
similar tensor structure to that of the dimension-six amplitude 
in Eq.~(\ref{eq:dim6}):
\begin{equation}
  \mathcal{M}_{\rm tree} = +\frac{g_s}{\sqrt{2}} \left| g_{r,\ell} \right|^2
  \bar u(q_1) T^c \gamma_{\nu} P_{R,L} v(q_2)
  \frac{ i \epsilon^{\mu \nu \alpha \beta} \epsilon^{*c}_{\mu} q_{3 \alpha}
    (q_1 + q_2)_{\beta} }
  {\left[ (-K/2 + q_1)^2 - M_{\widetilde q_{R,L}}^2 \right]
    \left[ (K/2 - q_2)^2 - M_{\widetilde q_{R,L}}^2 \right]}.
\end{equation}
Such a form will allow us to easily read off the interference term.  

The matrix element can be expanded in powers of $m_{\chi}^2/M_{\widetilde q}^2$
for $M_{\widetilde q} \gg m_{\chi}$,
yielding a leading $(1/M_{\widetilde q})^4$ behavior.  The leading term
comes from neglecting the $(-K/2 + q_1)^2$ and $(K/2 - q_2)^2$ terms in
the propagators compared to $M_{\widetilde q}^2$:
\begin{equation}
  \mathcal{M}_{\rm tree} \simeq 
  +\frac{g_s}{\sqrt{2}} \left| g_{r,\ell} \right|^2
  \bar u(q_1) T^c \gamma_{\nu} P_{R,L} v(q_2)
  \frac{ i \epsilon^{\mu \nu \alpha \beta} \epsilon^{*c}_{\mu} q_{3 \alpha}
    (q_1 + q_2)_{\beta} }
  {M_{\widetilde q_{R,L}}^4}.
\end{equation}
As a check, we can now compute the zero-velocity neutralino annihilation cross 
section from the tree-level $\chi\chi \to q \bar q g$ process, neglecting
the final-state quark mass.  Squaring
the matrix element, summing over final-state polarizations and colors and
averaging over initial-state polarizations, we obtain
\begin{equation}
  \frac{1}{4} \sum_{\rm pols} \left| \mathcal{M}_{\rm tree} \right|^2
  = 2 g_s^2 \left[ \frac{|g_r|^4}{M_{\widetilde q_R}^8}
    + \frac{|g_{\ell}|^4}{M_{\widetilde q_L}^8} \right]
  q_1 \cdot q_2 \left[ \left( q_1 \cdot q_3 \right)^2 + 
    \left( q_2 \cdot q_3 \right)^2 \right],
\end{equation}
for a single final-state quark flavor,
where we include the contributions from exchange of both $\widetilde q_L$
and $\widetilde q_R$.  Integrating over the phase space, we find
\begin{equation}
  v_{\rm rel} \sigma(\chi\chi \to q\bar q g) 
  = \frac{4}{15} \frac{m_{\chi}^6}{(4 \pi)^2}
  \left[ \frac{|g_r|^4}{M_{\widetilde q_R}^8} 
    + \frac{|g_{\ell}|^4}{M_{\widetilde q_L}^8} \right] \alpha_s
  \qquad\qquad ({\rm tree \ level}).
\end{equation}
This agrees with the result of Drees et al.~\cite{Dreesgg} when their
Eq.~(2.18) is expanded to extract the leading $1/M_{\widetilde q}^8$ 
dependence.

The cross section for the similar tree-level process 
$\chi \chi \to f \bar f \gamma$ was computed in Ref.~\cite{FOR} for the 
case of pure photino LSPs and degenerate left- and right-handed squarks,
$M_{\widetilde q_L} = M_{\widetilde q_R} \equiv M_{\widetilde q}$.
Replacing the color factor with the appropriate electric charge entails
$\alpha_s \rightarrow \alpha Q_f^2/4$.
The photino couplings are $g_{\ell} = -g_r = \sqrt{2} e Q_f$, so that
the couplings $|g_{r,\ell}|^4$ become
$|g_{r,\ell}|^4 \rightarrow 4 e^4 Q_f^4$.
Thus we obtain, for the tree-level (dimension-eight) 
$\chi\chi \to f \bar f \gamma$ with $m_f = 0$ and $v_{\rm rel} \to 0$,
\begin{equation}
  v_{\rm rel} \sigma(\chi\chi \to f \bar f \gamma) 
  = \frac{8}{15} \frac{m_{\chi}^6}{M_{\widetilde q}^8} 
  \alpha^3 Q_f^6
  \qquad\qquad ({\rm tree \ level}).
\end{equation}
Note that this result is a factor of two smaller than that of
Ref.~\cite{FOR}.

\subsection{Interference term}

We now combine the dimension-six and dimension-eight contributions.
While the dimension-six gluon-splitting amplitude in Eq.~(\ref{eq:dim6}) 
contains a purely vectorlike final-state quark current,
the dimension-eight amplitude contains both vectorlike and axial-vectorlike
quark currents:
\begin{eqnarray}
  \mathcal{M}_{\rm tree} &=& \frac{g_s}{\sqrt{2}} 
  i \epsilon^{\mu\nu\alpha\beta} \epsilon^{*c}_{\mu} q_{3\alpha}
  (q_1 + q_2)_{\beta}
  \nonumber \\ && \times
  \left\{ \frac{1}{2} \left[ \frac{|g_r|^2}{M_{\widetilde q_R}^4}
    + \frac{|g_{\ell}|^2}{M_{\widetilde q_L}^4} \right]
  \bar u(q_1) T^c \gamma_{\nu} v(q_2)
  + \frac{1}{2} \left[ \frac{|g_r|^2}{M_{\widetilde q_R}^4}
    - \frac{|g_{\ell}|^2}{M_{\widetilde g_L}^4} \right]
  \bar u(q_1) T^c \gamma_{\nu} \gamma_5 v(q_2) \right\}.
\end{eqnarray}
Only the vectorlike part of $\mathcal{M}_{\rm tree}$ will interfere
with the dimension-six amplitude.  Summing over final-state polarizations
and colors and averaging over initial-state polarizations, we obtain
the interference term,
\begin{equation}
  \frac{1}{4} \sum_{\rm pols} 2 \, {\rm Re} \left[ \mathcal{M}_{\rm tree} 
  \mathcal{M}_{\rm split}^* \right]
  = - 4 \alpha_s^2 
  \sum_{q^{\prime}} \left[ \frac{|g_r|^2}{M_{\widetilde q^{\prime}_R}^2}
    + \frac{|g_{\ell}|^2}{M_{\widetilde q^{\prime}_L}^2} \right]
  \sum_q \left[ \frac{|g_r|^2}{M_{\widetilde q_R}^4}
    + \frac{|g_{\ell}|^2}{M_{\widetilde q_L}^4} \right]
  \left[ (q_1 \cdot q_3)^2 + (q_2 \cdot q_3)^2 \right],
\end{equation}
where we have summed over the five light internal quarks $q^{\prime}$ in 
$\mathcal{M}_{\rm split}$ of Eq.~(\ref{eq:dim6}) and over the five light external quarks $q$.
Integrating over the phase space, we find the contribution to the cross 
section from the interference term,
\begin{equation}
  v_{\rm rel} \sigma_{\rm int} = - \frac{\alpha_s^2}{32 \pi^3} m_{\chi}^2
  \frac{2 m_{\chi}^2}{3}
  \sum_{q^{\prime}} \left[ \frac{|g_r|^2}{M_{\widetilde q^{\prime}_R}^2}
    + \frac{|g_{\ell}|^2}{M_{\widetilde q^{\prime}_L}^2} \right]
  \sum_q \left[ \frac{|g_r|^2}{M_{\widetilde q_R}^4}
    + \frac{|g_{\ell}|^2}{M_{\widetilde q_L}^4} \right].
\label{eq:intresult}
\end{equation}
We see that the form of this interference term is rather similar to that of
the dimension-six cross section given in Eq.~(\ref{eq:dim6xsec}).
In the special case of degenerate right- and left-handed squarks,
$M_{\widetilde q_R} = M_{\widetilde q_L} \equiv M_{\widetilde q}$,
we find that the interference term is related to the dimension-six cross
section by a multiplicative constant:\footnote{We disagree with the 
statement of Ref.~\cite{FOR} that the interference is zero
for photino-like neutralinos when $M_{\widetilde q_R} = M_{\widetilde q_L}$.
This conclusion was due to a spurious $\gamma_5$ in the loop amplitude matrix
element of Ref.~\cite{FOR}, which resulted in it interfering with the 
axial-vector part of the dimension-eight amplitude.}
\begin{equation}
  v_{\rm rel} \sigma_{\rm int} = v_{\rm rel} \sigma_{\rm dim6}
  \left[ - \frac{2}{3} \frac{m_{\chi}^2}{M_{\widetilde q}^2} \right].
\end{equation}

This result can be compared to the NLO corrections to 
$\chi\chi \to gg$~\cite{Barger05}; for $N_f = 5$ light flavors and taking the 
renormalization scale $\mu = 2 m_{\chi}$ we obtain
\begin{equation}
  v_{\rm rel}\sigma = v_{\rm rel} \sigma_{\rm dim6}
  \left[ 1 + \frac{221}{12} \frac{\alpha_s^{(5)}(2 m_{\chi})}{\pi}
    - \frac{2}{3} \frac{m_{\chi}^2}{M_{\widetilde q}^2} \right].
\end{equation}
The second term in the square brackets is the NLO correction 
from Ref.~\cite{Barger05} and 
the third term is our result for the interference term.  Note that for
$m_{\chi} \sim 100$ GeV the NLO correction is roughly 60\%; for, e.g., 
$M_{\widetilde q} \sim 2 m_{\chi}$, the interference term cancels off
about one quarter of the NLO correction.

Finally, we note that we have \emph{not} included a term of order 
$\alpha_s^2 m_{\chi}^4/M_{\widetilde q}^6$ from the expansion of the 
squark propagator in $\chi \chi \to gg$ in powers of $1/M_{\widetilde q}^2$.
This is because the loop integral is odd in the argument 
$(m_{\chi}^2/M_{\widetilde q}^2)$ with the next term in the series 
contributing to the cross section at order 
$\alpha_s^2 m_{\chi}^6/M_{\widetilde q}^8$, which is beyond our
present interest.

\section{Conclusions}

We have computed the cross section for neutralino annihilation
$\chi \chi \to q \bar q g$ at order $\alpha_s^2/M_{\widetilde q}^6$
arising from interference between the tree-level and loop-induced
processes, in the limit of zero relative neutralino velocity and ignoring
the masses of the five light quark flavors.  Our result for the total
cross section from this interference term is given in 
Eq.~(\ref{eq:intresult}).  We also provided the spin-averaged squares of
matrix elements for the dimension-six, dimension-eight, and interference
term contributions to $\chi\chi \to q \bar q g$ in terms of the final-state
particle momenta in order to facilitate their implementation into
Monte Carlos for the computation of indirect dark matter detection rates.

\begin{acknowledgments}
VB and GS were supported in part by the U.S.~Department of Energy
under grants DE-FG02-95ER40896 
and in part by the Wisconsin Alumni Research Foundation. 
W-YK was supported in part by the U.S.~Department of 
Energy under grant DE-FG05-84ER40173.
HEL was supported in part by the Natural Sciences and Engineering
Research Council of Canada.  VB thanks the Aspen Center for Physics for hospitality during the completion of this work.
\end{acknowledgments}

\appendix
\section{}

The matrix element for $\chi \chi \to gg$ with on-shell final-state
gluons and massless quarks in the loop can be expressed as an expansion in 
powers of $m_{\chi}^2/M_{\widetilde q}^2$ as follows:
\beq
    {\cal M}_{\rm loop} = \frac{\alpha_s}{2\sqrt{2} \pi}
    \sum_{q^{\prime}} \left[ 
      \frac{|g_r|^2}{M_{\widetilde q_R^{\prime}}^2} 
      F(m_{\chi}^2/M_{\widetilde q_R^{\prime}}^2)
      + \frac{|g_{\ell}|^2}{M_{\widetilde q_L^{\prime}}^2} 
      F(m_{\chi}^2/M_{\widetilde q_L^{\prime}}^2) \right]
    i \epsilon^{\mu\nu\alpha\beta} \epsilon^{*a}_{\mu}(k_1) 
    \epsilon^{*a}_{\nu}(k_2) k_{1\alpha} k_{2\beta},
    \label{eq:Mloopexp}
\eeq
where 
\beq
   F(a) = 1 + \frac{a^2}{9} + \frac{a^4}{25} + \cdots 
   + \frac{a^{2n}}{(2n+1)^2} + \cdots
\eeq
with $a = m_{\chi}^2/M_{\widetilde q}^2$, in agreement with 
Ref.~\cite{Dreesgg}.


\end{document}